\documentclass{article}
\usepackage{ijcai22}

\usepackage{times}

\usepackage{soul}
\usepackage{url}
\usepackage[hidelinks]{hyperref}
\usepackage[utf8]{inputenc}
\usepackage[small]{caption}
\usepackage{booktabs}
\usepackage{xcolor}

\usepackage{booktabs}
\pdfoutput=1
\usepackage{bbm}
\usepackage{mathptmx}
\usepackage{amssymb}
\usepackage{amsthm,amsfonts,bm}
\usepackage{amsmath,mathtools}
\usepackage{graphicx} 

\newtheorem{definition}{Definition}

\newtheorem{lemma}{Lemma}

\newtheorem{proposition}{Proposition}

\newtheorem{theorem}{Theorem}

\newtheorem{corollary}{Corollary}

\definecolor{ao(english)}{rgb}{0.0, 0.5, 0.0}
\definecolor{americanrose}{rgb}{1.0, 0.01, 0.24}
\definecolor{cerisepink}{rgb}{0.93, 0.23, 0.51}
\definecolor{darkorchid}{rgb}{0.6, 0.2, 0.8}
\definecolor{applegreen}{rgb}{0.55, 0.71, 0.0}
\definecolor{brightpink}{rgb}{1.0, 0.0, 0.5}
\definecolor{azure(colorwheel)}{rgb}{0.0, 0.5, 1.0}
\definecolor{blue-violet}{rgb}{0.54, 0.17, 0.89}
\definecolor{deepmagenta}{rgb}{0.8, 0.0, 0.8}
\definecolor{fashionfuchsia}{rgb}{0.96, 0.0, 0.63}
\definecolor{armygreen}{rgb}{0.29, 0.33, 0.13}
\definecolor{cobalt}{rgb}{0.0, 0.28, 0.67}
\definecolor{airforceblue}{rgb}{0.36, 0.54, 0.66}
\definecolor{amaranth}{rgb}{0.9, 0.17, 0.31}

\DeclareMathOperator {\marg}{\mathtt{marg}}

\title{Bayesian Promised Persuasion: Dynamic Forward-Looking Multiagent Delegation with Informational Burning}
\author{
Tao Zhang$^1$\footnote{Contact Author}\and
Quanyan Zhu$^1$\\
\affiliations
$^1$Electrical and Computer Engineering, New York University\\
\emails
\{tz636, qz494\}@nyu.edu
}

\begin{document}

\maketitle

\begin{abstract}

This work studies a dynamic mechanism design problem in which a principal delegates decision makings to a group of privately-informed agents without the monetary transfer or burning. We consider that the principal privately possesses complete knowledge about the state transitions and study how she can use her private observation to support the incentive compatibility of the delegation via informational burning, a process we refer to as the looking-forward persuasion. The delegation mechanism is formulated in which the agents form belief hierarchies due to the persuasion and play a dynamic Bayesian game. We propose a novel randomized mechanism, known as Bayesian promised delegation (BPD), in which the periodic incentive compatibility is guaranteed by persuasions and promises of future delegations. We show that the BPD can achieve the same optimal social welfare as the original mechanism in stationary Markov perfect Bayesian equilibria. A revelation-principle-like design regime is established to show that the persuasion with belief hierarchies can be fully characterized by correlating the randomization of the agents' local BPD mechanisms with the persuasion as a direct recommendation of the future promises.

\end{abstract}

\section{Introduction}

Building efficient rational multi-agent system is an important research agenda in Artificial Intelligence.
In many application domains, a system designer (\textit{principal}, she) aims to optimize the performance (e.g., social welfare or revenue) of a system in which multiple self-interested \textit{agents} actively behave.
The principal can use \textit{mechanism design} approaches to determine how agents should strategically interact with each other.
She can also influence the agents' decision makings by restricting their discretion (\textit{delegation}) or limiting their knowledge about the system (\textit{persuasion}). 
%
%
%

The key component of the system design problem is to provide \textit{incentives} that align the interests of participants.
Information asymmetry between the principal and the agents, however, imposes challenges for the incentive design.
Mechanism design literature studies two forms of information asymmetry: unobserved
preferences and unknown actions which leads to \textit{adverse selection} (see, e.g., \cite{myerson1981optimal,lobel2019dynamic}) and \textit{moral hazard} (see, e.g., \cite{bohren2019persistence,khalili2017designing}), respectively.

Monetary value has been long explored in designing incentive compatible (IC) mechanisms.
%
There are two forms of monetary value: \textit{transfer} and \textit{burning}.
Monetary transfer is also referred to as \textit{payment} that is paid by one party to the other. 
For example, in auctions, the winning bidder has to pay for the items to the auctioneer.
Hence, the money reduced from the bidder directly benefits the auctioneer.
Monetary burning, on the other hand, captures the setting when decisions are costly to one party but with no direct benefits to the other.
For example, the actions in delegation problems may incur a certain amount of wasteful expenditure (i.e., money burned) such as consumption of natural resources that may reduce the welfare of both the agents and the principal.
However, in certain scenarios, monetary value may be improper. 
For example, it is infeasible for organizations to use monetary incentive to efficiently allocate internal resources (e.g., vehicles, conference rooms).
In some cases, using monetary value may be controversial (e.g.,, medical or humanitarian-aid resource allocations) or constrained (incentive mechanism may violates the financial constraints.)
%
%
%

Information also plays an essential role in aligning the agents' behaviors with the principal's desire.
Bayesian persuasion (\cite{kamenica2011bayesian}) or information design studies how a principal can incentivize agents to behave in her favor through strategic information provision.
Instead of using monetary transfer or burning, we consider a process known as \textit{informational burning} to dynamically support the IC of the delegation design.
In particular, we consider an \textit{adverse selection} problem in a dynamic environment and study incentive compatible (IC) delegation mechanisms, in which a socially maximizing principal repeatedly delegates the decision makings to the agents by choosing an action menu, knowing that each agent will privately observe a payoff-relevant \textit{type} and then take an action from this menu.
The dynamic of the environment is driven by the transitions of the \textit{state of the world} (state) in a Markovian fashion where agents' current actions have persistence on future states.
The Markovian transition kernel of the state is parameterized by a variable known as \textit{shock}.
We assume that only the principal observes the shock and then controls what and how each agent gets to know about the shock---\textit{information burned}---in each period to influence the agents' decisions in each period.
In addition to the delegation rule, the principal also commits to a \textit{signaling rule} profile that performs \textit{Bayesian persuasion} to periodically inform the agents about the shock by sending \textit{signals} in each period.
The agents make current decisions by considering the future expected payoffs.
As a result, the principal's delegation design must dynamically capture the agents current and the \textit{planned} behaviors.
Since the shock realized in each period is ex-interim payoff relevant to the agents, the principal is able to influence each agent's current action choices by persuading the future.
We refer to such process as \textit{looking-forward persuasion} and the corresponding mechanism as the LFD mechanism.
%
%
%
%
%
%
%
%

%

We propose a novel model referred to as Bayesian Promised Delegation (BPD) which is a randomized mechanism composed of a signaling rule profile and a \textit{randomization} \textit{rule} profile.
The signaling rule performs the persuasion about the shock and the randomization rule jointly generates a pair of action menu and a promise of future expected payoff to each agent based on the current state and the belief hierarchies that are unavoidably induced by the persuasion.
By imposing incentive compatibility and promise-keeping constraints, we show that each social welfare that is achieved by an LFD mechanism can also be achieved by a BPD mechanism.
Furthermore, we establish a revelation-principle-like design regime to show that the informational burning can be fully characterized by a direct BPD in which the explicit formulations of the induced belief hierarchies is avoided by persuading the agents in terms of the recommendation of the promise.

\subsection{Related Work}
There is an extensive literature on mechanism design problems with transfer from static settings (see, e.g., \cite{myerson1981optimal}) to dynamic environments (see, e.g., \cite{pavan2014dynamic,lobel2019dynamic,zhang2021incentive}).
Monetary burning has been studied in general mechanism design problems (see, e.g., \cite{hartline2008optimal}) as well as delegation problems (see., e.g., \cite{amador2020money,ambrus2017delegation,Breig2016DelegationWC}).
There are also works studying mechanism designs without monetary incentives (see, e.g., \cite{procaccia2013approximate,johnson2014dynamic,guo2015dynamic,Breig2016DelegationWC,balseiro2019multiagent}).
\cite{Breig2016DelegationWC} has studied an optimal delegation problem when the principal and the agent cannot make payments to each other but are able to commit to actions in the future.
Their model uses the notion of continuation value that can be applied to dynamic delegation with and without monetary burning.
\cite{balseiro2019multiagent} has studied a resource allocation problem in which a principal repeated allocate a single resource in each period to one of multiple agents.
Their model does not consider the monetary transfer or burning; instead, the agents are incentivized to truthfully report their private information by promises or threat of future allocations.

Since the seminal work of \cite{kamenica2011bayesian}, there is a growing literature on Bayesian persuasion and information design (see, e.g., \cite{rayo2010optimal,ely2017beeps,bergemann2019information,hahn2020prophet,babichenko2020bayesian,celli2020private,mathevet2020information,zhang2021informational}).
This paper is closely related to the works studying \textit{direct} models of persuasion (or information design) by establishing design regimes similar to the Revelation Principle of the mechanism design.
In single-agent cases, \cite{kamenica2011bayesian} has shown that for every optimal persuasion model, there is a direct information structure that directly recommends the action to the receiver.
\cite{bergemann2016bayes} has considered multi-agent cases and shown that persuasion for a mixed-strategy Bayes Nash equilibrium has an equivalent Bayes correlated equilibrium model that performs direct action recommendation.
\section{Model and Problem Formulation}

\paragraph{Conventions.} For any measurable set $Y$, $\Delta(Y)$ denotes the set of probability measures over $Y$.
%
Any function defined on a measurable set is assumed to be measurable.
We use tilde to indicate the random variables; i.e., $y$ is a realization of the random variable $\tilde{y}$.
%
%
For a given expectation operator $\mathbb{E}[\cdot]$, $\mathbb{E}[F(\tilde{y}, w)]$ is the expectation taken over $Y$ while a realization $w$ is given, for any measurable function $F$.
Notation summary, proofs and remarks are provided in the online supplementary document.

We consider a discrete-time infinite horizon problem in a dynamic environment, where there is one \textit{principal} (she) and a finite number of \textit{agents} (he).
Time is indexed by $t=0, 1,2,\dots$ and agents are indexed by $i\in \mathcal{N}\equiv[n]$ with $1\leq n < \infty$.
The principal repeatedly delegates decision makings to the agents.
In each period $t$, each agent $i$ takes an \textit{action} $a_{i,t}\in A_{i,t}\subseteq A$, where $A$ is a compact set of actions and $A_{i,t}\subseteq A$ is an \textit{action menu} determined by the principal in period $t$.
The dynamic of the environment is characterized by three notions of \textit{information}.
At the beginning of each period $t$, a \textit{state} $s_{t}\in S$ is drawn.
Based on $s_{t}$, each agent $i$ observes his \textit{type} $\theta_{i,t}\in \Theta$ which is drawn according to the distribution $d^{\Theta}_{i}(\cdot|s_{t})\in \Delta(\Theta_{i})$, for all $i\in \mathcal{N}$.
Both the state $s_{t}$ and the type $\theta_{i,t}$ are the payoff-relevant information of each agent $i$ in each period $t$.
%
In addition, the environment generates a \textit{shock}, $x_{t}\in X$, which is drawn according to the distribution $d^{X}(\cdot|s_{t})\in \Delta(X)$.
We consider that $S$, $\Theta$, and $X$ are compact sets of states, types, and shocks, respectively.
%
%

The state of the environment evolves in a Markovian fashion. Formally, let $h_{t}=(s_{0}, a_{0}, x_{0}, \dots, s_{t-1}, a_{t-1}, x_{t-1})$ denote the \textit{history} up to period $t$.
The state in period $t$ depends on history $h_{t}$ only through the state of period $t-1$.
If the state in period $t-1$ is $s_{t-1}$, the agents take joint action $a_{t-1}=(a_{i,t-1})_{i\in\mathcal{N}}$, and $x_{t-1}$ is the shock in period $t-1$, then, period-$t$ state is distributed as
\begin{equation*}
    \begin{aligned}
    \tilde{s}_{t}\sim \kappa(\cdot|s_{t-1}, a_{t-1}, x_{t-1}),
    \end{aligned}
\end{equation*}
where the function $\kappa: S\times A_{t} \times X\mapsto \Delta(S)$ is the \textit{transition kernel} with $A_{t}=\prod_{i\in\mathcal{N}}A_{i,t}$.
The initial distribution of the state is given as $\kappa_{0}(\cdot)\in \Delta(S)$.

We assume that the environment is \textit{two-sided information-asymmetric.} 
Specifically, the state $s_{t}$ is publicly observed by the principal and all the agents.
However, the type $\theta_{i,t}$ is the private information of each agent $i$ that is unobserved by the principal and other agents while the shock $x_{t}$ is only observed by the principal.
On the one hand, the principal has imperfect information about the agents' decision makings of choosing actions because she does not know the agents' type profile, $\theta_{t}=(\theta_{i,t})_{i\in\mathcal{N}}$.
On the other hand, each agent $i$ has imperfect information about other agents' decision makings because he does not know others' types, $\theta_{-i,t}$; additionally, he has uncertainty regarding the distribution of the next-period state due to the unobservability of the current shock, $x_{t}$.
By knowing the shock, the principal designs how and what each agent should know about the shocks in addition to the action menu.
Specifically, the principal informs each agent $i$ about the shock $x_{t}$ by privately sending agent $i$ a \textit{signal} $\omega_{i,t}\in \Omega_{i}$ where $\Omega_{i}$ is a compact set of signals, for $i\in\mathcal{N}$.
We refer to the tuple $\mathcal{E}\equiv<\mathcal{N}, S, A, X, \Theta, \kappa, d^{X}, \{d^{\Theta}_{i}\}_{i\in \mathcal{N}}>$ with the aforementioned two-sided information asymmetry as the \textit{environment model}.
%

\subsection{Dynamic Delegation Mechanism}


We consider that the principal designs stationary delegation mechanisms in the dynamic environment.
Hence, we suppress the time index for the elements in the environment models, unless otherwise stated.
Specifically, the principal uses a \textit{menu function} $\lambda_{i}(\cdot):\Theta_{i,t} \mapsto A$ to determine an action menu $A_{i}$ for each agent $i$ when the current state is $s$. 
That is,
\begin{equation}\label{eq:action_menu}
    A_{i}= \big\{a_{i}\in A: a_{i} = \lambda_{i}(\theta_{i}), \theta_{i}\in\Theta_{i}\big\}.
\end{equation}
%
Let $\Lambda_{i}$, denote a compact set of menu functions the principal can choose from for each agent $i$, for $i\in\mathcal{N}$. 
We restrict attention to \textit{contingent delegation} setting, in which each menu function $\lambda_{i}$ is generated according to a stationary mixed-strategy \textit{delegation rule}, $\sigma_{i}:S \mapsto \Delta(\Lambda_{i})$, such that 
$\sigma_{i}(\lambda_{i}|s)\in[0,1]$ gives the probability of specifying a menu function $\lambda_{i}$ for agent $i$ when the state is $s$.
Here, the action menu $A_{i}$ specified for each agent $i$ depends on the current state $s$ only through the randomized generation of the menu function $\lambda_{i}$ by $\sigma_{i}(\cdot|s)$.

Additionally, the principal uses a stationary mixed-strategy \textit{signaling rule} $\phi_{i}:S\times X \mapsto \Delta(\Omega_{i})$ to select a signal $\omega_{i}\in \Omega_{i}$ for each agent $i$.
The signaling rule profile performs Bayesian persuasion for the agents.
In this work, we restrict attention to Markovian mechanisms in which each menu function and each signaling rule only take into consideration the current relevant information and are independent of histories.
We refer to such delegation mechanism with looking-forward persuasion as an LFD mechanism, denoted by $D\equiv<\sigma, \phi>$ where $\sigma=(\sigma_{i})_{i\in\mathcal{N}}$ and $\phi=(\phi_{i})_{i\in\mathcal{N}}$.

We assume that both the principal and the agents share the same discount factor denoted by $\delta\in(0,1)$.
In every period $t$, each agent $i$ chooses an action $a_{i,t}$ from the menu $A_{i}$ (uniquely determined by $\lambda_{i}$) to maximize his period-$t$ expected payoff including period-$t$ one-stage utility and discounted sum of total utility of the future starting from period $t+1$.
The objective of the principal is to maximize the ex-ante social welfare, i.e., the ex-ante expected discounted sum of utilities of all the agents, by designing an incentive compatible LFD mechanism to restrict agents' discretion (through delegation) and their additional information about the unobserved shocks (through persuasion).
\subsection{Belief Model}\label{sec:belief_system}

In this work, we consider that each agent is a Bayesian player. That is, each agent forms \textit{beliefs} about the unobserved shock and his opponents' types and signals.
We assume that the generation of each agent $i$'s type is from \textit{a move by Nature} (i.e., a \textit{Harsanyi's type} \cite{harsanyi1967games}) with $d^{\Theta}\equiv(d^{\Theta}_{i})_{i\in\mathcal{N}}$ as common priors.
That is, a type $\theta_{i}$ is a full description of agent $i$'s beliefs about the data of the game, beliefs about the beliefs of the his opponents about the data of the game and about his own beliefs, etc. (see, e.g., \cite{zamir2020bayesian}).
Hence, the information asymmetry raised from the private types is \textit{information imperfectness.}

However, the information asymmetry due to the unobservability of the shock and other agents' signals is \textit{information incompleteness,} which unavoidably induces agents' interactive reasoning about the beliefs.
Specifically, since the principal's $\phi$ is publicly known, each agent $i$ forms a \textit{posterior belief} $\mu_{i}(\cdot|s, \omega_{i})\in\Delta(X\times \Omega_{-i})$ about the shock $x$ and other agents' signals $\omega_{-i}$, using $\phi$ and $d^{X}$ according to the Bayes' law.
Since each agent $i$'s period-$t$'s choice of action takes into consideration of the future, the posterior belief $\mu_{i}$ is payoff-relevant.
As other agents' actions $a_{-i}$ are payoff-relevant to agent $i$, so are their posterior beliefs $\mu_{-i}$.
As a result, each agent $i$ needs to form beliefs about other agents' posterior beliefs.
For the same reason, each agent $i$ has to form a belief about other agents' beliefs of his beliefs of $\mu_{-i}$, and so on.
Therefore, this information incompleteness leads to a \textit{belief hierarchy} for each agent in each period.

Formally, a belief hierarchy is an infinite sequence of beliefs, $b_{i}\equiv(b^{[1]}_{i}, b^{[2]}_{i}, \dots, b^{[k]}_{i}, \dots)$.
Here, $b^{[1]}_{i} = \marg_{X} \mu_{i}(\cdot|s, \omega_{i})\in \Delta(X)$ is the first-order belief about the unobserved shock $x$, given the state $s$ and the signal $\omega_{i}$.
Since every agent $j\neq i$ has a first-order belief $b^{[1]}_{j}$, agent $i$ uses $(s, \omega_{i})$ to form a belief about $b^{[1]}_{j}$.
Hence, each agent $i$ forms a second-order belief $b^{[2]}_{i}\in \Delta(X\times \Delta(X)^{n-1})$ about $b^{[1]}_{-i}$.
For the same reason, the $k$th-order belief $b^{[k]}_{i}$ is a belief about the shock $x$ and others' $k-1$th-order beliefs $b^{[k-1]}_{-i}$, and so on.
Therefore, in each period $t$, each signal $\omega_{i}$ induces a belief hierarchy for each agent $i$ given any state $s$; i.e., there is a correspondence $\Xi_{i}(\cdot;s)$ such that $b_{i} = \Xi_{i}(\omega_{i};s)$. Let $b = \Xi(\omega;s)=(\Xi_{i}(\omega_{i};s))_{i\in\mathcal{N}}$.
Since the principal knows the signals sent to the agents, she knows each agent $i$'s belief hierarchy $b_{i}$ in every period, for all $i\in\mathcal{N}$.

Let $\mathring{B}^{[k]}_{i}$ denote the set of $k$th-order beliefs of agent $i$, for $i\in\mathcal{N}$.
Let $\bar{B}^{[k]}_{i}=\bar{B}^{[k-1]}_{i}\times \mathring{B}^{[k]}_{i}$ denote the set of $k$-level belief hierarchies with $\bar{B}^{[1]}_{i} = \Delta(X)$.
In each period, a belief hierarchy $b_{i}$ is \textit{coherent} (\cite{brandenburger1993hierarchies}) if $b^{[k]}_{i}$, for any $k>1$, coincides with all beliefs of lower order, $\{b^{[k']}_{i}\}_{k'=1}^{k-1}$; i.e., $b^{[k-1]}_{i} = \marg\limits_{\mathring{B}^{[k-1]}_{i}} b^{[k]}_{i}$ for all $i\in\mathcal{N}$, $t\geq0$, $k>1$.
%
%
In a coherent belief hierarchy, any event in the space of the $k$th-order beliefs, $\mathring{B}^{[k]}_{i}$, must have the same marginal probability in every $k'$th-order beliefs, for all $k'>k$.
Let $B^{[k]}_{i}\subseteq \bar{B}^{[k]}_{i}$ denote the set of coherent belief hierarchies of order $k$ of agent $i$, for all $i\in\mathcal{N}$, $k\geq 1$, with $B^{[1]}_{i} = \bar{B}^{[1]}_{i}$, $B_{i} = B^{[\infty]}_{i}$.
From the conherency, the projection of $B^{[k+1]}_{i}$ on $\bar{B}^{[k]}_{i}$ is $B^{[k]}_{i}$ (see, \cite{zamir2020bayesian}).
\cite{brandenburger1993hierarchies} has shown that there exists a \textit{homemorphism} $\Gamma_{i}(\cdot;s): \Omega \mapsto \Delta(X\times B_{-i})$ such that $\Gamma_{i}(x, b_{-i}|b_{i}, s)\in[0,1]$ gives agent $i$ a probability of an event that the shock is $x$ and other agents' (coherent) belief hierarchies are $b_{-i}$, when the state is $s$ and agent $i$'s belief hierarchy is $b_{i}$.
We assume that $\Gamma_{i}(\cdot;s)$ (with $\Gamma^{X}(\cdot|b_{i};s)=\marg\limits_{X}\Gamma_{i}(\cdot|b_{i}; s)$) is given for every $s\in S$, $i\in\mathcal{N}$, and it is publicly known.
Given the state $s$ and the belief hierarchy $b_{i}$, we denote the transition kernel perceived by agent $i$ by $\hat{\kappa}(s'|s, b_{i}, a) \equiv \int_{x} \kappa(s'|s, x, a)\Gamma^{X}_{i}(d x|b_{i}; s)$.
We assume that \textit{(i)} for all $(s, b_{i})\in S\times B_{i}$ and all $a\in A^{n}$, $\hat{\kappa}(\cdot|s, b_{i}, a)$ is absolutely continuous and \textit{(ii)} for all $(s, b_{i})\in S\times B_{i}$, the mapping $a\mapsto \hat{\kappa}(\cdot|s, b_{i}, a)$ is norm-continuous.

Define the \textit{belief hierarchy distribution} (belief distribution) induced by the principal's signaling rule as, for any $s\in S$, 
\begin{equation*}
    \begin{aligned}
    Z(b|s)\equiv \int\limits_{x\in X} \phi(\{\omega:\Xi(\omega;s)=b\}|s, x)d^{X}(d x|s).
    \end{aligned}
\end{equation*}
We let $Z_{i}(\cdot|s)\in \Delta(B_{i})$ denote the marginal distribution for each $i\in\mathcal{N}$.

Following \cite{mathevet2020information}, we establish the following three conditions that is sufficient and necessary for the existence of a signaling rule profile that induces a belief distribution $Z(\cdot|s)\in \Delta(B)$, for any $s\in S$.
\textit{(i)} the mechanism should hhave a \textit{common prior} $p(\cdot|s)\in \Delta(X\times B)$ such that, for all $i\in\mathcal{N}$, $s\in S$, $p(x, b|s) = \Gamma_{i}(x, b_{-i}|b_{i};s)p(b_{i}|s)$ where $p(b_{i}|s) = \int_{x\in X, b_{-i}\in B_{-i} } p(x, b_{-i}, b_{i})$.
\textit{(ii)} The belief distribution induced by the signaling rule is \textit{consistent}; i.e., for all $s\in S$, $Z(\cdot|s) = \marg\limits_{B}p(\cdot|s)$.
\textit{(iii)} The belief distribution induced by the signaling rule is \textit{Bayes' plausible}; i.e., for all $s\in S$, $\int\limits_{b_{i}\in B_{i}}\Gamma^{X}_{i}(\cdot|b_{i}; s)Z_{i}(d b_{i}|s) = d^{X}(\cdot|s)$.
We use $T=<p, Z, \Gamma, B>$ to denote the \textit{belief model} induced by the signaling rule profile $\phi$.
We refer to $T$ that satisfies \textit{(i)}-\textit{(iii)} as \textit{regular belief model.}
Given $T$, we expand the information structure $<\kappa, S>$ used in delegation rule profile to $\{<\kappa, S>, <Z, B>\}$ such that each $\sigma_{i}(\cdot|s, b_{i})\in\Delta(\Lambda_{i})$ for all $i\in\mathcal{N}$.

\subsection{Dynamic Bayesian Game}

Given any LFD mechanism $<\sigma, \phi>$ with the belief model $T$, the agents play a dynamic Bayesian game, in which each agent $i$ takes an action from the action menu $A_{i}$ in each period and receives a single-period utility.
Define each agent $i$'s \textit{utility function} as $u_{i}:S\times \Theta_{i} \times A \mapsto \mathbb{R}$ such that agent $i$ receives one-period utility $u_{i}(s, \theta_{i}, a)$ when the state is $s$, his type is $\theta_{i}$, and the agents take $a$.
We denote the underlying dynamic Bayesian game in the mechanism $<\sigma, \phi>$ as a tuple $M[\sigma, \phi]=<\mathcal{N}, S, A, \Theta, T, \{u_{i}\}_{i\in\mathcal{N}}>$.
We consider that in each period $t$, each agent $i$ uses a pure-strategy \textit{Markov policy} $\pi_{i,t}: s\times \Theta_{i} \times B_{i} \mapsto A_{i}$ to selects an action $\pi_{i,t}(s_{t}, \theta_{i,t}, b_{i,t})\in A_{i}$. 
We say that that a policy profile $\{\pi_{i,t}\}_{t\geq 0}$ is \textit{obedient} if agent $i$ selects an action $a_{i,t}$ according to his type $\theta_{i,t}$ in each period $t$;
i.e., $\pi_{i,t}(s_{t}, \theta_{i,t}, b_{i,t})=\lambda_{i}(\theta_{i,t})$, for all $s_{t}\in S$, $\theta_{i,t}\in\Theta_{i}$, $b_{i,t}\in B_{i}$, $t\geq 0$.
%

Following the revelation principle, it is without loss of generality to focus on mechanisms where agents are obedient.
In particular, we consider \textit{Markov perfect Bayesian equilibria} (MPBE) in obedient policies, in which each agent believes with probability $1$ that other agents are obedient.
To incentivize the obedience, we impose \textit{periodic ex-interim incentive compatibility} (PIC) constraints to ensure that under the delegation mechanism, obedience is each agent's best response to other agents' obedience.

Due to the Ionescu Tulcea theorem (see, e.g., \cite{hernandez2012discrete}), the initial distribution $\kappa_{0}$ on $S$, the transition kernel $\kappa$, conditional probability distributions $<d^{\Theta}, d^{X}>$, the delegation mechanism rule profiles $<\sigma, \phi>$ with the belief model $T$, and the agents' policy profile $\pi=(($ $\pi_{i,t}$ $)_{i\in\mathcal{N}})_{t=0}^{\infty}$ uniquely define a probability measure $P^{\sigma, \phi}_{\pi}$ on $(S\times\Theta\times X\times B \times A)^{\infty}_{0}$.
In addition, given any $(s_{t}, \theta_{i,t},$ $b_{i,t}, a_{i,t})$, each agent $i$ perceives a unique probability measure $P^{\sigma, \phi}_{\pi}[s_{t}, \theta_{i,t}, b_{i,t}, a_{i,t}]$ on $\Theta_{-i}\times B_{-i}\times(S\times\Theta\times X$ $\times B \times A)_{t+1}^{\infty}$.
The expectation operators with respect to $P^{\sigma, \phi}_{\pi}$ and $P^{\sigma, \phi}_{\pi}[s_{t}, \theta_{i,t}, b_{i,t}]$ are denoted by $\mathbf{E}^{\sigma, \phi}_{\pi}$ and $\mathbb{E}^{\sigma, \phi}_{\pi}[\cdot|s_{t}, \theta_{i,t}, b_{i,t}]$, respectively.
Then, we define the \textit{ex-interim expected payoff to-go} (payoff to-go) of agent $i$ as: 
\begin{equation}\label{eq:interim_expected_payoff}
    \begin{aligned}
    &J^{\sigma, \phi}_{i}(a_{i,t}, s_{t}, \theta_{i,t}, b_{i,t};\pi)\equiv (1-\delta)\mathbb{E}^{\sigma,\phi}_{\pi}\Big[ u_{i}(s_{t}, \theta_{i,t}, a_{i,t}, \tilde{a}_{i,t})\\
    &+ \sum\limits_{\ell = t+1} \delta^{\ell-t}u_{i}(\tilde{s}_{\ell}, \tilde{\theta}_{i,\ell}, \tilde{a}_{\ell}) \Big| s_{t}, \theta_{i,t}, b_{i,t}, a_{i,t}\Big].
    \end{aligned}
\end{equation}
%
%
When any agent $i$'s policy is obedient, we omit it in the notation of the payoff to-go function (e.g., $J^{\sigma, \phi}_{i}(\lambda_{i}(\theta_{i,t}), s_{t}, \theta_{i,t}, b_{i,t})$ and $J^{\sigma, \phi}_{i}(a_{i,t}, s_{t}, \theta_{i,t}, b_{i,t};\pi_{i})$ if $\pi$ and $\pi_{-i}$ are obedient, respectively.)
%
%
Then, we define the PIC constraints as follows: for all $i\in\mathcal{N}$, $t\geq 0$, $s_{t}\in S$, $\theta_{i,t}\in \Theta_{i}$, $b_{i,t}\in B_{i}$, $a_{i,t}\in A_{i}$,
\begin{equation}\tag{$\mathtt{PIC}_{i,t}$}\label{eq:PIC_original}
    \begin{aligned}
    J^{\sigma, \phi}_{i}(\lambda_{i}(\theta_{i,t}), s_{t}, \theta_{i,t}, b_{i,t})\geq J^{\sigma, \phi}_{i,t}(a_{i,t}, s_{t}, \theta_{i,t}, b_{i,t};\pi_{i}).
    \end{aligned}
\end{equation}
We say that an LFD mechanism $<\sigma, \phi>$ is PIC if $<\sigma, \phi>$ satisfies (\ref{eq:PIC_original}), for all $i\in\mathcal{N}$, $t\geq 0$.
If the principal implements such $<\sigma, \phi>$, then each agent $i$, believing with probability $1$ that others are obedient, being obedient in every period is a MPBE.
As is standard, when there are multiple equilibria, tie-breaking rule is in the principal's favor.

Under the mechanism $<\sigma,\phi>$ with $T$, each agent $i$'s \textit{ex-ante expected payoff} is defined as $J_{i}(\sigma, \phi, \pi)\equiv \mathbf{E}^{\sigma, \phi}_{\pi}\Big[ \sum\limits_{t=0}^{\infty}\delta^{t}u_{i}(\tilde{s}_{t}, \tilde{\theta}_{i,t}, \tilde{a}_{t}) \Big]$.
When $\pi$ is obedient, we write $J_{i}(\sigma, \phi)=J_{i}(\sigma, \phi, \pi)$.
We refer to the vector $(J_{i}(\sigma, \phi, \pi))_{i\in\mathcal{N}}$ as the principal's \textit{target} of the mechanism.
We assume that the every target $J_{i}$ is bounded for all $i\in\mathcal{N}$.
We define the set of \textit{attainable targets} of the principal as follows:
\begin{equation}
    \begin{aligned}
    V\equiv \Big\{v\in \mathbb{R}^{n}\Big| v_{i} = J_{i}(\sigma, \phi), i\in\mathcal{N},\text{ for a PIC } <\sigma, \phi>   \Big\}.
    \end{aligned}
\end{equation}
Hence, for any vector $v=(v_{i})_{i\in\mathcal{N}}\in \mathtt{G}$, there exists a PIC LFD mechanism $<\sigma, \phi>$ under which each agent $i$ is incentivized to adopt the obedient policy and obtains ex-ante expected payoff as $v_{i}$, for all $i\in\mathcal{N}$.
Since the principal aims to maximize the ex-ante social welfare which is the ex-ante expected discounted sum of all the agents' utilities, an optimal PIC delegation mechanism $<\sigma^{*}, \phi^{*}>$ has a corresponding $v^{*}\in \mathcal{V}$ such that $\sum\limits_{i\in\mathcal{N}}J_{i}(\sigma^{*}, \phi^{*}) = \sum\limits_{i\in\mathcal{N}}v^{*}_{i}=\max\limits_{v\in G}\sum\limits_{i\in\mathcal{N}} v_{i}$.


\section{Dynamic Bayesian Promised Delegation Model}

In this section, we propose the model of \textit{Bayesian promised delegation} (BPD) mechanism in the same environment $\mathcal{E}$.
A BPD is a randomized mechanism consisting of a signaling rule profile $\phi^{\natural}=(\phi^{\natural}_{i})_{i\in\mathcal{N}}$ and a \textit{randomization rules} profile $\psi=(\psi_{i})_{i\in\mathcal{N}}$.
Similar to $\phi$ in the original delegation mechanism, the principal uses each $\phi^{\natural}_{i}$ to inform each agent $i$ about the realized shock in every period.
As shown in Sec. \ref{sec:belief_system}, the profile $\phi^{\natural}$ induces a belief model $T^{\natural}=<p, Z, \Gamma, B>$.
We first define \textit{attainable state value}. Then, we introduce the notion of \textit{stage rules} and formally define the BPD model.
Finally, we establish a revelation principle to obtain a \textit{direct BPD} mechanism in which the signaling profile is fully characterized by the direct randomization rule profile.


\subsection{Attainable State Value}

In the LFD mechanism $<\sigma, \phi>$, each agent $i$ periodically decides whether his choice of action $a_{i,t}$ can maximize his current payoff to-go, $J^{\sigma, \phi}_{i}(a_{i,t}, s_{t}, \theta_{i,t}, b_{i,t};\pi_{i})$ which is composed of current expected one-period utility $\hat{u}_{i}(s_{t}, \theta_{i,t}, a_{i,t}) \equiv \mathbb{E}^{\tilde{\theta}_{-i,t}}\big[u_{i}(s_{t}, \theta_{i,t}, a_{i,t}, \lambda_{-i}(\tilde{\theta}_{-i,t}))|s_{t}\big]$
and a \textit{continuing value}, $c^{\sigma,\phi}_{i}(s_{t}, b_{i,t}, a_{i,t};\pi^{t+1:}_{i}) \equiv (1-\delta)\mathbb{E}^{\sigma,\phi}_{\pi^{t+1:}_{i}}\Big[\sum\limits_{\ell=t+1}^{\infty}\delta^{\ell-t}u_{i}(\tilde{s}_{\ell}, \tilde{\theta}_{i,\ell}, \tilde{a}_{\ell})\Big| s_{t}, b_{i,t} ,a_{i,t}\Big]$,
%
where $\pi^{t+1:}_{i} = (\pi_{i,\ell})_{\ell=t+1}^{\infty}$.
When $\pi^{t+1:}_{i}$ is obedient, we write 
$c^{\sigma,\phi}_{i}(s_{t},  b_{i,t}, \lambda_{i}(\theta_{i,t}))=c^{\sigma,\phi}_{i}(s_{t},  b_{i,t}, a_{i,t};\pi^{t+1:}_{i})$.
%
%

\begin{lemma}\label{lemma:one_shot_deviation}
Fix an LFD mechanism $<\sigma, \phi>$ with belief model $T$. 
Then, the delegation mechanism is PIC if and only if, for all $i\in\mathcal{N}$, $s\in S$, $\theta_{i}, \theta'_{i}\in \Theta_{i}$, $b_{i}\in B_{i}$, 
\begin{equation*}
    \begin{aligned}
    (1-\delta)\hat{u}_{i}(s, &\theta_{i}, \lambda_{i}(\theta_{i})) + c^{\sigma,\phi}_{i}(s, b_{i}, \lambda_{i}(\theta_{i}))\\
    &\geq (1-\delta)\hat{u}_{i}(s, \theta_{i}, \lambda_{i}(\theta'_{i})) + c^{\sigma,\phi}_{i}(s, b_{i}, \lambda_{i}(\theta'_{i})).
    \end{aligned}
\end{equation*}
\end{lemma}

Lemma \ref{lemma:one_shot_deviation} establishes a \textit{one-shot deviation principle} which directly follows the subgame perfectness of Markov perfect Bayesian equilibria and we omit the proof here. 
Although the delegation mechanism is stationary, each agent's deviation from obedience can be arbitrarily nonstationary. That is, in each period $t$, every agent $i$ can use current policy $\pi_{i,t}$ and plan future policies $\pi^{t+1:}_{i}$.
Lemma \ref{lemma:one_shot_deviation} implies that if the delegation mechanism is incentive compatible when each agent might deviate from obedience only in current period, then it is also incentive compatible for agents' arbitrary deviations.
%

%
Define the \textit{state value function} $g^{\sigma, \phi}_{i}(s) = \mathbb{E}^{ \sigma, \phi}\Big[\sum\limits_{t=0}^{\infty}\delta^{t}u_{i}(\tilde{s}_{t}, \tilde{\theta}_{i,t}, \lambda(\tilde{\theta}_{t}))\Big|s\Big]$, assuming that agents are obedient.
%
%
%
It is straightforward to obtain that in any PIC LFD mechanism $<\sigma, \phi>$, the state value function $g^{\sigma, \phi}_{i}$ is uniquely defined by the recursion: $g^{\sigma, \phi}_{i}(s) = \mathbb{E}^{\sigma, \phi}\Big[(1-\delta)u_{i}(s, \tilde{\theta}_{i}, \lambda(\tilde{\theta})) + \delta\int\limits_{s'}g^{\sigma,\phi}_{i}(s')\hat{\kappa}(d s'|s, \tilde{b}_{i}, \lambda(\tilde{\theta})) \Big|s\Big]$.
%
Let $g^{\sigma, \phi}(\cdot)=(g^{\sigma,\phi}_{i}(\cdot))\in\mathbb{R}^{n}$ denote the vector of agents' state value functions.
Then we define the set of \textit{attainable state value} functions:
\begin{equation}
    \begin{aligned}
    G\equiv \Big\{g \Big| g_{i}(\cdot) = g^{\sigma, \phi}_{i}(\cdot), i\in\mathcal{N},\text{ for a PIC } <\sigma,\phi>\Big\}.
    \end{aligned}
\end{equation}

\begin{corollary}\label{corollary:equal_g_v}
For every attainable target $v\in\mathcal{V}$, there is a $g\in G$ such that $v_{i} = \mathbf{E}^{\sigma, \phi}[g_{i}(\tilde{s})]$, for all $i\in\mathcal{N}$.
\end{corollary}

\subsection{Stage Rules and BPD Model}

Lemma \ref{lemma:one_shot_deviation} implies that the expected next-period obedient payoff-to-go, characterized by $g^{\sigma, \phi}$, constitutes a sufficient statistic for determining whether an LFD mechanism is incentive compatible in the current period.
The key design principle for our BPD mechanisms is to allow the principal to provide a \textit{promise} of the future delegations in addition to the action menus.
Define a pure-strategy \textit{stage menu function} profile, denoted by $\lambda^{\natural}=(\lambda^{\natural}_{i})_{i\in\mathcal{N}}\in \Lambda$, and a \textit{stage promise rule}, denoted by $\rho=(\rho_{i})_{i\in\mathcal{N}}\in P = \prod_{i\in\mathcal{N}} P_{i}$ where $P_{i}\equiv\{\rho_{i}:S\mapsto \mathbb{R}\}$ is a compact set of promise rules for agent $i$, for all $i\in\mathcal{N}$.
Here, each stage menu function $\lambda^{\natural}_{i}:\Theta_{i}\mapsto A$ defines an action menu $A_{i}$ in the same way as the menu function $\lambda_{i}$ in (\ref{eq:action_menu}).
When the current period state is $s$, agent $i$'s belief hierarchy is $b_{i}$,
each $\rho_{i}:S\mapsto \mathbb{R}$ specifies a \textit{promise} function such that we obtain the \textit{expected promised value} of the next period for agent $i$, $L_{i}(s, b_{i}, a; \rho_{i})\equiv  \int\limits_{s'}\rho_{i}(s')\hat{\kappa}(d s'|s, b_{i}, a)$.
%
%
%

Formally, a BPD mechanism model is defined by $\mathtt{BPD}[\phi^{\natural},\psi,P]\equiv<\phi^{\natural}, \psi, T^{\natural}, P, \Lambda >$.
The principal uses the signaling rule profile $\phi^{\natural}$ to send signals to the agents such that a belief hierarchy profile $b$ is formed.
Each randomization rule $\psi_{i}(\cdot|s, b_{i}) \mapsto \Delta(\Lambda_{i} \times P_{i} )$ specifies the probability of privately generating a pair of stage rules, $<\lambda^{\natural}, \rho_{i}>$, for agent $i$ when the state is $s$ and the agent's belief hierarchy is $b_{i}$.
%

The Ionescu Tulcea theorem implies that the initial distribution $\kappa_{0}$ on $S$, the transition kernel $\kappa$, conditional probability distributions $<d^{\Theta}, d^{X}>$, the profiles $<\phi^{\natural}, \psi>$, and the induced belief model $T^{\natural}$ uniquely define a probability measure over $(S\times\Theta\times X\times B \times \Lambda \times P \times A)^{\infty}_{0}$.
We use $\mathbb{E}^{\psi}[\cdot|\cdot]$ and $\mathbf{E}^{\psi}[\cdot]$, respectively, denote the ex-interim (for any specific conditions) and the ex-ante expectation operators.
Define the one-stage \textit{ex-post expected payoff} of agent $i$ as $R_{i}(s, \theta_{i}, b_{i}, a ; \rho_{i}) \equiv (1-\delta)u_{i}(s, \theta_{i}, a) +\delta L_{i}(s, b_{i}, a; \rho_{i})$.
%
%
Given $\mathtt{BPD}[\phi^{\natural},\psi,P]$, define the one-stage \textit{ex-interim expected payoff} of agent $i$ as $\bar{R}_{i}(s, \theta_{i}, b_{i}, a_{i}; \rho_{i}|\psi_{-i})\equiv \mathbb{E}^{\psi}\Big[R_{i}(s, \theta_{i}, b_{i}, a_{i}, \tilde{\lambda}_{-i}(\tilde{\theta}_{-i}); \rho_{i})|s, b_{i}\Big]$.
%
%
%

\begin{definition}[Bayesian Incentive Compatibility]
A mechanism $\mathtt{BPD}[\phi^{\natural},\psi,P]$ is Bayesian incentive compatible (BIC) if 
for all $i\in\mathcal{N}$, $s\in S$, $\theta_{i}\in\Theta_{i}$, $b_{i}\in B_{i}$, $\lambda^{\natural}_{i}\in \Lambda_{i}$ and $\rho_{i}\in P_{i}$ with $\psi_{i}(\lambda^{\natural}_{i}, \rho_{i}|s, b_{i})>0$, $a'_{i}\in A_{i}$,
\begin{equation}\tag{$\mathtt{BIC}_{i}$}\label{eq:BPD_IC}
    \begin{aligned}
    \bar{R}_{i}(s, \theta_{i}, b_{i}, \lambda^{\natural}_{i}(\theta_{i}); \rho_{i}|\psi_{-i}) \geq \bar{R}_{i}(s, \theta_{i},b_{i}, a'_{i}; \rho_{i}|\psi_{-i}),
    \end{aligned}
\end{equation}
while each agent beliefs with probability $1$ that other agents are obedient.
\end{definition}

The inequality displayed in (\ref{eq:BPD_IC}) is referred to as the BIC constraint for agent $i$.
It ensures that agent $i$, after observing $(s, \theta_{i})$ and the stage rules $(\lambda_{i},\rho_{i})$ and forming $b_{i}$ (induced by a signal sent by $\phi^{\natural}_{i}$), finds that it is better off for him to be obedient while believing with probability $1$ that others are obedient.
Given the obedience, the profile $z(\cdot)=\{z_{i}(\cdot)\}_{i\in\mathcal{N}}$ where each $z_{i}(\cdot|s, b_{i}) = \int\limits_{\theta_{i}}\marg\limits_{\Lambda_{i}}\psi_{i}(\lambda_{i}(\theta_{i}),\cdot|s, b_{i})d^{\Theta}_{i}(\theta_{i}|s)$ constitutes a Bayes Nash equilibrium.

%

We also require the model $\mathtt{BPD}[\phi^{\natural},\psi,P]$ to satisfy the following \textit{Bayesian promise keeping} (BK) constraints: for all $i\in\mathcal{N}$, $s\in S$, $g_{i}\in P_{i}$,
\begin{equation}\tag{$\mathtt{BK}_{i}[g_{i}]$}\label{eq:BPD_BPK}
    \begin{aligned}
    g_{i}(s) = \mathbb{E}^{\psi}\Big[ \bar{R}_{i}(s, \tilde{\theta}_{i}, \tilde{b}_{i}, \tilde{\lambda}^{\natural}_{i}(\tilde{\theta}_{i}); \tilde{\rho}_{i}|\psi_{-i})\Big|s\Big].
    \end{aligned}
\end{equation}
%
%
Let $\mathtt{BK}[g]\equiv \{\mathtt{BK}_{i}[g_{i}]\}_{i\in\mathcal{N}}$.
Specifically, $\mathtt{BK}[g]$ requires the principal to promise a current state-value function profile $g$ by promising future state-value function profile through the randomization rule $\psi$.
We refer to a BPD mechanism satisfying BIC and $\mathtt{BK}[g]$ as BIC-$\mathtt{BK}[g]$ mechanism.
Based on the constraints $\mathtt{BIC}$ and $\mathtt{BK}[g]$, we define the following set: for any compact subset $\hat{P}\subseteq P$,
\begin{equation}
    \begin{aligned}
    H[\hat{P}] \equiv \Big\{ g\in P\;|\; \exists  \mathtt{BPD}[\phi^{\natural},\psi,\hat{P}], \text{ s.t., } \mathtt{BIC}, \mathtt{BK}[g] \Big\}.
    \end{aligned}
\end{equation}

\begin{proposition}\label{prop:fixed_point_set}
The set of attainable state-value functions, $G$, satisfies $H[G]=G$.
\end{proposition}

Proposition \ref{prop:fixed_point_set} implies that set of attainable state value functions of the RD mechanisms is a \textit{``fixed point"} of $H[\cdot]$ and for every $g\in G$ there is a BIC and BK $\mathtt{BPD}[\phi^{\natural},\psi,\hat{P}]$ that can achieve the same $g$.
According to Corollary \ref{corollary:equal_g_v}, we conclude that any social welfare that can be achieved by a RD mechanism can also be achieved by some BPD mechanism.

\subsection{Correlated Bayesian Promised Delegation}

In this section, we introduce the \textit{direct} BPD (DPD) mechanisms in which the signal set coincides with the set of promise rules; i.e., $\Omega_{i} = P_{i}$, for all $i\in\mathcal{N}$.
Formally, a DPD mechanism is defined by the model $\mathtt{DPD}[\eta, P]\equiv<\eta, P, \Lambda>$, in which $\eta:S\times X \mapsto \Delta(\Lambda\times P)$ is a \textit{direct randomization rule}.
Note that $\eta$ is not a profile.
In each period, the principal generates a profile of stage rule pairs $(<\lambda_{i}, \rho_{i}>)_{i\in\mathcal{N}}$ according to $\eta(\cdot|s, x)$ based on the state $s$ and the shock $x$.
Each $<\lambda_{i}, \rho_{i}>$ is privately sent to each agent $i$.
Based on $<\lambda_{i}, \rho_{i}>$ and his type $\theta_{i}$, agent $i$ chooses an action from the menu defined by $\lambda_{i}$.

Let $\mathbf{E}^{\eta}[\cdot]$ and $\mathbb{E}^{\eta}[\cdot|\cdot]$ denote the corresponding expectation operators.
Given any $(s,x)\in S\times X$, define the belief distribution induced by the signaling rule profile $\phi^{\natural}$ as: $Z^{\natural}(b|s,x) \equiv \phi^{\natural}\big(\big\{\omega:\Xi(\omega;s) = b\big\} \big)$.
%
%
%
With a slight abuse of notation, let $\hat{R}_{i}(s, \theta_{i}, x, a; \rho_{i})\equiv (1-\delta)u_{i}(s, \theta_{i}, a) +\delta \int\limits_{s'}\rho_{i}(s')\kappa(ds'|s, x, a)$.
Then, define $m_{i}(\theta_{i}, a_{i};\lambda_{i},\rho_{i}|s,x) \equiv \mathbb{E}^{\eta}\big[ \hat{R}_{i}(s, \theta_{i}, x, a_{i}, \tilde{\lambda}_{-i}(\tilde{\theta}_{-i}); \rho_{i})\big|s,x \big]$.
%

\begin{definition}[Bayesian Correlated Incentive Compatibility]
A BPD mechanism $\eta$ is Bayesian correlated incentive compatible (BCIC) if for each $i\in\mathcal{N}$, $s\in S$, $\theta_{i}\in \Theta_{i}$, $\lambda_{i}\in\Lambda_{i}$, $\rho_{i}\in P_{i}$,
\begin{equation}\label{eq:BCIC_def}
    \begin{aligned}
    \mathbb{E}^{\eta}\Big[m_{i}(\theta_{i}, \lambda_{i}(\theta_{i}); \lambda_{i}, \rho_{i}|s, \tilde{x})\Big|s\Big] \geq \mathbb{E}^{\eta}\Big[m_{i}(\theta_{i}, a'_{i}; \lambda_{i}, \rho_{i}|s, \tilde{x})\Big|s\Big],
    \end{aligned}
\end{equation}
for all $a'_{i}\in A_{i}$.
\end{definition}

The inequality displayed in (\ref{eq:BCIC_def}) is referred to as the BCIC constraint for agent $i$.
This constraint guarantees that agent $i$, after observing $(s, \theta_{i})$ and stage rules $(\lambda_{i}, \rho_{i})$, finds that it is optimal to be obedient when he believes with probability $1$ that other agents are obedient.
Given the obedience, the distribution $\hat{z}(\cdot|s,x)=\int\limits_{\theta} \marg\limits_{\Lambda}\eta(\lambda(\theta), \rho|s, x)\prod\limits_{i\in\mathcal{N}} d^{\Theta}_{i}(\theta_{i})$ constitute a Bayes correlated equilibrium. 
Besides the BCIC constraints, we require the mechanism $\mathtt{DPD}[\eta, P]\equiv<\eta, P, \Lambda>$ to satisfy
following \textit{Bayesian correlated promise keeping} (BCK) constraints: for all $i\in\mathcal{N}$, $s\in S$, $g_{i}\in P_{i}$,
\begin{equation}\tag{$\mathtt{BCK}_{i}[g_{i}]$}\label{eq:DPD_BCK}
    \begin{aligned}
    g_{i}(s) = \mathbb{E}^{\eta}\Big[ m_{i}(\tilde{\theta}_{i}, \tilde{\lambda}_{i}(\tilde{\theta}_{i}); \tilde{\lambda}_{i}, \tilde{\rho}_{i}|s, \tilde{x})\Big|s\Big].
    \end{aligned}
\end{equation}
Let $\mathtt{BCK}[g]\equiv\{\mathtt{BCK}_{i}[g_{i}]\}_{i\in\mathcal{N}}$.
We refer to a DPD mechanism satisfying BIC and $\mathtt{BCK}[g]$ as BCIC-$\mathtt{BCK}[g]$ mechanism.
We say that a randomization profile $\psi$ \textit{induces} a direct randomization rule $\eta$ if, for all $s\in S$, $x\in X$, $\lambda\in \Lambda$, $\rho\in G$, 
\begin{equation*}
    \begin{aligned}
    \eta(\lambda, \rho|s, x) \equiv\int\limits_{b}  \Big(\prod_{i\in\mathcal{N}} \psi_{i}(\lambda_{i}, \rho_{i}|s, b_{i})\Big)Z^{\natural}(db|s,x).
    \end{aligned}
\end{equation*}

\begin{theorem}\label{thm:revelation_DPD}
For any PIC LFD mechanism $<\sigma, \phi>$ that achieves an attainable $g\in G$, there exists a BCIC-$\mathtt{BCK}[g]$ DPD mechanism $\mathtt{DPD}[\eta, G]$ that achieves the same $g$ if there exists a BIC-$\mathtt{BK}[g]$ BPD mechanism $\mathtt{BPD}[\phi^{\natural}, \psi, G]$ that induces $\mathtt{DPD}[\eta, G]$.
%
%
\end{theorem}

Following Corollary \ref{corollary:equal_g_v}, Theorem \ref{thm:revelation_DPD} implies that for every attainable target $v\in V$ that can be achieved by a PIC LFD mechanism $<\sigma, \phi>$, there exists a BCIC DPD mechanism that can achieve $v$ if there exists a BIC BPD mechanism that induces this DPD mechanism.
Therefore, an equivalence of social welfare is obtained.
Theorem \ref{thm:revelation_DPD} establishes our version of \textit{revelation principle} for the looking-forward persuasion.
In particular, the persuasion with information structure $<\phi, \Omega>$ which induces belief hierarchies for the agents can be fully characterized by the randomized mechanism DPD in which the randomization of the agents' individual stage rules are correlated with the recommendation of the future promises as the direct persuasion.
As a result, the explicit formulation of belief hierarchies is avoided.

\section{Conclusion}

In this work, we have studied a dynamic delegation mechanism design problem with informational burning that is supported by a looking-forward persuasion. 
%
%
We have proposed a novel randomized mechanism known as Bayesian promised delegation (BPD) model in addition to the signaling rule that jointly generates an action menu and a promise of future to each agent based on the state and agent's belief hierarchy induced by the persuasion.
We have shown that by imposing the incentive compatibility and the promise keeping constraints, BPD mechanism can achieve a same social welfare as the original LFD mechanism.
A revelation principle is obtained to show that the delegation with informational burning can be fully characterized by a direct BPD in which we avoid explicit formulations of the belief hierarchies by persuading each agent through direct recommendation of the promise.
These results contribute as foundations for the algorithmic analysis of the dynamic mechanism design which is our next step.


\bibliographystyle{named}
\bibliography{ijcai22}

\end{document}